\documentclass[aps,prl,toolkits,twocolumn]{revtex4}

\usepackage{epsfig,amssymb,amsfonts}

\makeatletter

\renewcommand{\@makefntext}[1]
{\parindent=1em\noindent\hbox to 1.8em{\hss$^{\@thefnmark}$}#1}
\renewcommand{\@footnotemark}{\hbox{\mathsurround=0pt$^{\@thefnmark}$}}

\makeatother

\begin{document}
\title{Comment on ``Is the spectrum of highly excited mesons purely
coulombian?"}

\maketitle

The high-lying spectrum of light baryons and mesons reveals a high
degree of degeneracy - there appears  parity doubling \cite{G1,G2,JPS,G3} 
as well as
some degeneracy with respect to the hadron spin \cite{G3,G4,A,K}. 
In a recent letter \cite{GON} a fit  within the nonrelativistic
potential approach has been done to some  recently found high-lying meson
states \cite{AN,B} that exhibit a pattern of degeneracy. 
The following explanation for degeneracy is suggested. At large
separations between valence quarks in a highly excited meson the confining
interaction is screened by  string breaking and the system is governed
by the color Coulomb interaction only. This provides a required degeneracy,
like in the nonrelativistic Hydrogen atom. In particular,
the states with different orbital angular momenta $L$ and radial quantum
number $n_r$ are degenerate within the band with the principal quantum
number $ \sim n_r +L$.

It is unclear, however, how could it be that the confining color-electric
interaction is completely screened by the string breaking at large
distances between valence quarks, while at the same time the color-Coulomb
interaction is still operating. Second, within the Coulomb problem  the bound
state spectrum
becomes very dense in the upper part, close to the end of the descrete spectrum, 
a property that is not observed in the
meson spectrum. Third, the strong coupling constant,
required by the fit is very large, $\alpha_s = 9.4$, which implies
that the system becomes relativistic (in contrast to the Hydrogen atom).
The color-magnetic spin-spin interaction  accompanying the Coulomb
interaction in  nonrelativistic systems, which is not taken into
account in \cite{GON}, induces hyperfine splittings $\sim \alpha_s^4$,
that would totally ruin the degeneracy.

The nonrelativistic picture of constituent quarks
does not contain the notion of chirality.
However, once the relativistic nature of  quarks is taken into
account via the Dirac formalism, practically massless ultrarelativistic
quarks in  highly-excited hadrons become chiral and the chiral symmetry is  
restored,  a natural microscopic
explanation of parity doubling \cite{G3,WG,L}.
A degeneracy between  states with different spins $J$ can be obtained
if one assumes that in the semiclassical regime there appears a
principal quantum number $\sim n_r + J$ \cite{GN}. Note that such a
principal quantum number is consistent with  Lorentz invariance.
Assuming the existence of a
principal quantum number $\sim n_r + L$ \cite{A2,K,SV,F} would
imply a separate conservation of quark spins $S$ and their angular momenta
$L$ (i.e., of the angular momentum of the gluonic field) in excited hadrons.
It is hard to reconcile such an assumption with  Lorentz invariance.

A key experimental question is to find chiral (parity) partners to
the high spin states, that are presently missing in refs. \cite{AN,B}
or to reliably establish their absence. Note that such high spin
parity doublets are well seen in the established nucleon spectrum.
For a recent study of the delta spectrum see ref. \cite{bonn}.

\noindent
The author acknowledges support of the Austrian Science Fund through
Grant No. P19168-N16.

{ Leonid Ya. Glozman} \\ \\
 {Institute for
Physics, University of Graz, Universit\"atsplatz 5, A-8010 Graz,
Austria}


\begin{thebibliography}{99}

\bibitem{G1} L. Ya. Glozman, Phys. Lett. {\bf B 475}, 329 (2000);
T. D. Cohen, L. Ya. Glozman, Phys. Rev. {\bf D 65}, 016006 (2002).
\bibitem{G2} L. Ya. Glozman, Phys. Lett. {\bf B 539}, 257 (2002);
L. Ya. Glozman, Phys. Lett. {\bf B 587}, 69 (2004).
\bibitem{JPS} R. L. Jaffe, D. Pirjol, A. Scardicchio, Phys. Rep.
{\bf 435}, 157 (2006).
\bibitem{G3}  L. Ya. Glozman, Phys. Rep. {\bf 444},1 (2007).
\bibitem{G4} L. Ya. Glozman, Phys. Lett. {\bf B 541}, 115 (2002). 
\bibitem{A} S. S. Afonin, Eur. Phys. J. {\bf A 29}, 327 (2006).
\bibitem{K}  E. Klempt, A. Zaitsev, Phys. Rep. {\bf 454}, 1 (2007).
\bibitem{GON} El Houssine Mezoir and P. Gonzales, arXiv:0810.5651 [hep-ph],
accepted in PRL.
\bibitem{AN} A. V. Anisovich et al, Phys. Lett. {\bf B 491}, 47 (2000);
{\bf B 517}, 261 (2001); {\bf B 542}, 8 (2002);{\bf B 542}, 19 (2002).
\bibitem{B} D. V. Bugg, Phys. Rep. {\bf 397}, 257 (2004).
\bibitem{WG} R. F. Wagenbrunn, L. Ya. Glozman,
Phys. Rev. {\bf D 75}, 036007 (2007).
\bibitem{L} A. V. Nefediev, J. E. F. Ribeiro, A. P. Szczepaniak,
JETP Lett. {\bf 87}, 271 (2008); F. J. Llanes-Estrada et al,
arXiv:0810.4462 [hep-ph].
\bibitem{GN} L. Ya. Glozman, A. V. Nefediev, Phys. Rev. {\bf D 76},
096004 (2007).
\bibitem{A2} S. S. Afonin, Phys. Rev. {\bf C 76}, 015202 (2007).
\bibitem{SV} M. Shifman, A. Vainshtein, Phys. Rev. {\bf D 77}, 034002 (2008).
\bibitem{F} H. Forkel, M. Beyer, T. Frederico, JHEP {\bf 0707}, 077 (2007).
\bibitem{bonn} I. Horn et al, arXiv:0711.1138 [nucl-ex].
\end{thebibliography}
\end{document}